# Sorting Algorithms with Restrictions[1]

## H. Aslanyan
### Yerevan State University
1, A. Manoukyan street, 375049 Yerevan

## 1. Introduction

Sorting is one of the most used and well investigated algorithmic problem [1]. Traditional postulation supposes the sorting data archived, and the elementary operation as comparisons of two numbers. In a view of appearance of new processors and applied problems with data streams, sorting changed its face. This changes and generalizations are the subject of investigation in the research below.

## 2. General notes, sorting of binary information

Let it is given a sequence of numbers $a_1,...a_n$. An elementary operation $(a_i : a_j)$ compares $a_i$ and $a_j$. Sorting is the reordering of $a_1,...a_n$ with increasing; - the known boundary of required number of operations $S(n)$ is known as

$$\lceil \log_2 n! \rceil \leq S(n) \leq \sum_{k=2}^{n} \left\lceil \log_2 \frac{3}{4} k \right\rceil$$

Roughly, the order of given expression is $n \log_2 n$. When the input sequence is ordered and we compare pairs of neighbours, we use only $n-1$ comparisons. The essential property of this class of algorithms is the way of selection of current pair of elements to be compared. There are other algorithms which don't use the previous steps. These are special sorting schemes where input is any sequence $a_1,...a_n$ and output is its ordered counterpart. The scheme itself is fixed, consisting of 2-comparators linked each to other in a special way. If $\overline{S}(n)$ is the required number of elements of a scheme, then $\overline{S}(n) \geq S(n)$ and the known estimation is as $\overline{S}(n) = O(n(\log_2 n)^2)$. Below a generalization of schemes from 2 to $k$-comparators will be brought.
One more notion is:

**Theorem 1(the 0-1 principle).**

If an $n$-scheme sorts all the $2^n$ $0-1$-sequences of length $n$, then it sorts any numeric sequences of length $n$.

## 3. $(n, k)$ Sorting Scheme

$(n, k)$ sorting scheme receives an n-length input and sorts k elements in one step. Let us denote the scheme size by $S(n, k)$. The simplest $(n,3)$ scheme looks like:

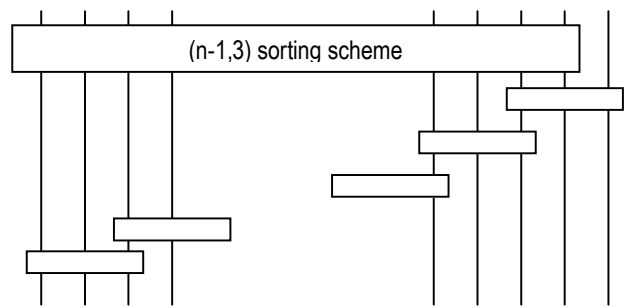

---

[1] The research is supported partly by INTAS: 04-77-7173 project, http://www.intas.be

The complexity approximation is $n^2/4$. Similarly might be constructed an $(n,k)$ scheme of complexity

$$\sum_{p=k}^{n}\frac{p-1}{k-1} = \frac{1}{k-1}\left(\sum_{p=1}^{n}(p-1) - \sum_{p=1}^{k-1}(p-1)\right) = \frac{n(n-1) - (k-1)(k-2)}{2(k-1)}.$$

This complexity might be reduced.

Let us formulate some postulations:

1. $S(n, 2n/3) = 3$
2. $S(n, 7n/12) \leq 4$
3. $S(n, n/2) = 5$

## 4. Parallel Sorting

Let's consider an algorithm, which is allowed during each step to perform in parallel arbitrary number of comparisons of pairs of elements, with the only restriction – each element occurs at most in one comparison. Let $L(n)$ be complexity of the algorithm for sorting $n$ elements. The following theorem takes place:

**Theorem 2**  $\quad L(n) \leq \dfrac{\log_2 n \cdot (1 + \log_2 n)}{2}$

The proof uses the following property of merge sort algorithm:

**Lemma** Let $A$ be a decreasing sequence of $n$ elements and $B$ - an increasing sequence of $m$ elements and $m \geq n$. Then it is possible to merge $A$ and $B$ and obtain a sorted sequence of $m+n$ elements by $\log(m+n)$ operations.

*We skip the proof of the lemma here.*

The table below presents the operations when we consequently partition the input sequence:

| Length of sorted subsequence | Number of subsequences | Number of operations at this stage |
|---|---|---|
| 1 | n | 0=0 |
| 2 | n/2 | 1=1 |
| 4 | n/4 | log2*2=2 |
| ... | ... | ... |
| N | 1 | log(2*n/2)=log(n) |

## 5. References

1. Д. Кнут, Искусство программирования, Том 3, Сортировка и поиск, второе издание, 2001, 832 стр.
2. Ռ.Ն Տոնոյան, Կոմբինատոր ալգորիթմներ, ԵՊՀ: